A top-down approach for desynchronization in an ensemble of coupled oscillators

Ming Luo[1] and Yongjun Wu[2]

1. School of Aerospace Engineering and Applied Mechanics, Tongji University, Shanghai, China 200092

2. School of Naval Architecture, Ocean and Civil Engineering, Shanghai Jiao Tong University, Shanghai, China, 200240

A universal approach is proposed for suppression of collective synchrony in a large population of interacting rhythmic units. We demonstrate that provided that the internal coupling is weak, stabilization of overall oscillations with vanishing stimulation leads to desynchronization in a large ensemble of coupled oscillators, without altering significantly the essential nature of each constituent oscillator. We expect our findings to be a starting point for the issue of destroying undesired synchronization, e. g. desynchronization techniques for deep brain stimulation for neurological diseases characterized by pathological neural synchronization.

Coupled nonlinear oscillators are abundant in biology, physics, and chemical reaction systems [1, 2, 3]. They exhibit various collective dynamic behaviors, among which synchronization is of the greatest importance. It exists ubiquitously in nature. Examples include pacemaker cells in the heart [2], glycolytic synchrony in yeast cell suspensions [4], and synchronously flashing fireflies [5], etc. Synchronization techniques are also widely used in modern techniques, such as arrays of lasers [6], superconducting junctions [7], and gene networks [8], etc. However, on some occasions unexpected synchronization is pernicious. An illustrative example is the crowd synchrony which induced the high-amplitude lateral sway of the London Millennium Bridge on its opening day [9]. In neurological diseases, such as Parkinson's diseases [10], the brain functioning is severely impaired by excessively synchronized firing of thousands of neurons. Therefore desynchronization in a large population of coupled oscillators is of crucial scientific and practical significance [11, 12].

In recently years, owing mainly to the significant clinical needs for mild stimulation to eliminate harmful pathological neural synchronization in networks of oscillatory neurons [13], desynchronization in ensembles of interacting units is attracting considerable interest [12]. Several novel methods have been proposed [14, 15, 16, 17]. Desynchronization effects of feedback, especially time-delayed feedback, have been extensively studied [15]. It have been demonstrated both theoretically and experimentally that feedback in the form of linear or nonlinear functions of the mean activities of a population suppresses coupling-induced synchronization [15, 18]. However, the basic question, how such external feedback changes internal synchronization, is not yet well understood till now. It is generally accepted that desynchronization is a "bottom-up" process, namely feedback signal restores the natural frequency of each oscillator, and thereby destroy the synchrony between individual oscillations so that they do not sum up coherently to a large mean field [15]. Most importantly, there is a lack of a general theory to deal with unwanted conditions in various different contexts [12].

In this manuscript, we first examine feedback effects on synchronization in two coupled oscillators, studying the mechanism by which the coherence is changed. A general approach is then presented for desynchronization in large populations of coupled oscillators. Its potential applications are illustrated in neuroscience with a network of coupled chaotic neural oscillators. Our findings may also shed light on the control of complex systems [19].

We consider a pair of coupled Landau-Stuart oscillators [20]

$$\dot{z}_1 = (1 + i\omega_1 - |z_1|^2)z_1 + \frac{\varepsilon}{2}(z_2 - z_1)$$
$$\dot{z}_2 = (1 + i\omega_2 - |z_2|^2)z_2 + \frac{\varepsilon}{2}(z_1 - z_2)$$
(1)

along with a linear mean field feedback ([15], with zero time-delay)

$$u(t) = -gZ(t)$$
(2)

which is applied equally to the two oscillators. $z_{1,2}$ are complex numbers which represent the states of the two oscillators at time $t$, $\omega_{1,2}$ are the natural frequencies, $\varepsilon \geq 0$ is the coupling strength, the mean field $Z(t) = (z_1 + z_2)/2$ represents the average behavior of two oscillators, and $g$ is feedback gain.

Let $z_1 = r_1 e^{i\theta_1}$, $z_2 = r_2 e^{i\theta_2}$, then the closed-loop system of (1) and (2) is of the following form

$$\dot{r}_1 = \left(1 - \frac{\varepsilon + g}{2} - r_1^2\right)r_1 + \frac{\varepsilon - g}{2}r_2 \cos\phi$$

$$\dot{r}_2 = \left(1 - \frac{\varepsilon + g}{2} - r_2^2\right)r_2 + \frac{\varepsilon - g}{2}r_1 \cos\phi \tag{3}$$

$$\dot{\phi} = \Delta - \frac{\varepsilon - g}{2}\left(\frac{r_1}{r_2} + \frac{r_2}{r_1}\right)\sin\phi$$

where $\phi = \theta_2 - \theta_1$ is the phase difference, and $\Delta = \omega_2 - \omega_1$ is the frequency mismatch. We assume that $\Delta \geq 0$. For the symmetry of the system (3), we consider the stable symmetric solutions of (3) with $r_1 = r_2 = \rho$, which satisfy [20]

$$\dot{\rho} = \left(1 - \frac{\varepsilon + g}{2} - \rho^2\right)\rho + \frac{\varepsilon - g}{2}\rho \cos\phi$$

$$\dot{\phi} = \Delta - (\varepsilon - g)\sin\phi \tag{4}$$

System (4) has no fixed points if $|g - \varepsilon| < \Delta$. For $|g - \varepsilon| > \Delta$, it has two fixed points. When $g < \varepsilon$, the two fixed points are given by

$$\rho_\pm^2 = 1 - \frac{\varepsilon + g}{2} \pm \frac{\sqrt{(\varepsilon - g)^2 - \Delta^2}}{2} \tag{5}$$

$$\phi_+ = \arcsin(\Delta/(\varepsilon - g)), \qquad \phi_- = \pi - \arcsin(\Delta/(\varepsilon - g))$$

When $g > \varepsilon$, the fixed points are

$$\rho_\pm^2 = 1 - \frac{\varepsilon + g}{2} \pm \frac{\sqrt{(\varepsilon - g)^2 - \Delta^2}}{2} \tag{6}$$

$$\phi_+ = \pi + \arcsin(\Delta/(\varepsilon - g)), \qquad \phi_- = 2\pi - \arcsin(\Delta/(\varepsilon - g))$$

The sign "+" in (5) and (6) represents stable fixed point and the "-" sign to unstable fixed point. The stable fixed point corresponds to the phase-locked state of system (1). The long-term behavior of the closed-loop system (1) and (2) in $g - \Delta$ space and $g - \varepsilon$ space are shown in Fig. 1a and Fig. 1b, respectively. Bifurcation curves are determined by $\varepsilon - g = -\Delta$ and $\varepsilon - g = \Delta$. Contrary to previous results [15], even the simple feedback is capable of desynchronization, though the phase drift or incoherent region (yellow regions) persists only for a narrow range of feedback gain $g$. Moreover, it is clear from equation (4) that the two oscillators exhibit rather complex behaviors than oscillations with their own natural frequencies in the incoherent region.

To examine how external feedback engineers internal coupling-induced synchronization, we study the evolution of the phase difference, the amplitude of the mean field and of individual oscillator with the feedback gain. According to equations (5) and (6), we obtain

$$\left.\begin{aligned}\rho^2 &= 1 - \frac{\varepsilon+g}{2} + \frac{\sqrt{(\varepsilon-g)^2 - \Delta^2}}{2} \\ \phi &= \arcsin(\Delta/(\varepsilon-g)), \qquad R = \frac{\rho}{2}\sqrt{2+2\cos\phi}\end{aligned}\right\}, \quad g < \varepsilon$$

$$\left.\begin{aligned}\rho^2 &= 1 - \frac{\varepsilon+g}{2} - \frac{\sqrt{(\varepsilon-g)^2 - \Delta^2}}{2} \\ \phi &= \pi + \arcsin(\Delta/(\varepsilon-g)), \qquad R = \frac{\rho}{2}\sqrt{2+2\cos\phi}\end{aligned}\right\}, \quad g > \varepsilon$$

(7)

where $R = |Z(t)|$ is the amplitude of the mean field. When $g < \varepsilon$, as $\varepsilon - g \to \Delta$, the individual amplitudes $r_1 = r_2 = \rho \to \sqrt{(2-\varepsilon+g)/2}$, the mean field amplitude $R \to \sqrt{2-(\varepsilon+g)}/2$, and the phase difference $\phi \to \pi/2$. When $g > \varepsilon$, as $g \to +\infty$, $\rho \to \sqrt{1-\varepsilon}$, $R \to 0$, and $\phi \to \pi$. An example is shown Fig. 2. We see that with increasing feedback gain $g$, the amplitude of the mean field $R$ approaches to zero, while the phase difference moves towards $\pi$. The two oscillators reach a state of anti-phase synchronization.

From the control point of view, the effects of external feedback can be explained as follows. Consider the dynamical system that governs the evolution of the mean field $Z(t)$. It has a fixed point at the origin (e. g. in the case of oscillation death [21]). The negative state feedback (2) stabilizes the mean field to this fixed point [22]. Under strong enough feedback, the mean field system converges to the origin with $R \to 0$. The coherence (characterized by the phase difference $\phi$) between the two oscillators is rearranged so that their oscillations sum up to the resultant zero mean field. Thus, we draw a conclusion, opposite to that of previous studies [15], that feedback control of synchronization is a "top-down" process. External feedback controls the mean field and thereby regulates synchronization between individual oscillators.

The next question is what is the impact of such feedback on synchronization process in a large ensemble of coupled oscillators. Therefore, we consider an ensemble of all-to-all coupled Landau-Stuart equations [21, 23]

$$\dot{z}_j = (1 + i\omega_j - |z_j|^2)z_j + \frac{\varepsilon}{N}\sum_{k=1}^{N}(z_k - z_j) \qquad (8)$$

where $j = 1, \cdots, N$. Obviously, the system governing the dynamics of the mean field $Z(t) = \frac{1}{N}\sum_{j=1}^{N} z_j$ also has a fixed point at the origin. A simple linear negative state feedback $u(t)$

$$u = -gZ(t) \qquad (9)$$

is added to the right hand of equation (8) to stabilize the complex mean field system [22].

In the case of large population, individual oscillation almost makes no contribution to the mean field $Z(t)$, i.e., $N \to \infty$, $z_j/N \to 0$. Thus, $\frac{1}{N}\sum_{k=1}^{N} z_k = \frac{1}{N}\sum_{k=1,k\neq j}^{N} z_k + \frac{z_j}{N} \approx \frac{1}{N}\sum_{k=1,k\neq j}^{N} z_k$. The closed-loop system (8) and (9) is approximated by

$$\dot{z}_j = (1 - \varepsilon + i\omega_j - |z_j|^2)z_j + \frac{\varepsilon}{N}\sum_{k=1,k\neq j}^{N} z_k - \frac{g}{N}\sum_{k=1,k\neq j}^{N} z_k \quad (10)$$

For this reason, the last two terms can be regarded as external stimulation to the *j*th oscillator. Furthermore, when the feedback gain $g$ is large enough, the collective mean field variations collapse to noise-level or zero amplitude ($Z(t) \to 0$). Under these conditions, the last two terms in equation (10) approach to zero and can be neglected. Equation (10) is further approximated by

$$\dot{z}_j = (1 - \varepsilon + i\omega_j - |z_j|^2)z_j \quad (11)$$

We should note that these conditions (referred to as large size effects) do not stand in the case of small sets. Equation (11) indicates that the ensemble (8) are decoupled by the feedback with large enough gain $g$. For weak coupling ($\varepsilon < 1$), all the elements oscillate incoherently with individual radii $\sqrt{1-\varepsilon}$ and individual natural frequencies $\omega_j$ as if they are uncoupled.

A numerical example of an ensemble ($N = 500$) is shown in Fig. 3. In the coupling- and feedback-free regime, the mean field only has small fluctuations of order $o(N^{-1/2})$ [23] (Fig. 3a, blue line, $t < 500$) due to the incoherent individual oscillations (Fig. 3b). The coupling ($\varepsilon = 0.5$) results in a rather large mean field (Fig. 3a, blue line, $500 < t < 1,000$). The feedback stabilizes the mean field (Fig. 3a, blue line, $1000 < t < 1,500$), and hence desynchronizes the ensemble (Fig. 3c). Note that once the mean field is suppressed, the feedback signals approach to zero (Fig. 3a, red line). However, the individual radii are changed from $\sqrt{1.0}$ to $\sqrt{1-\varepsilon} \approx 0.7071$ (Fig. 3c).

To demonstrate how to apply the approach and how it works in practical applications, we consider a realistic model of collective rhythmical activities in a population of Hindmarsh-Rose neurons [24] with time-delayed coupling

$$\begin{aligned}
\dot{x}_j &= y_j - x_j^3 + 3x_j^2 - z_j + I_j + \varepsilon X(t-\tau_0); \\
\dot{y}_j &= 1 - 5x_j^2 - y_j; \\
\dot{z}_j &= 0.006[4(x_j + 1.56) - z_j];
\end{aligned} \quad (13)$$

where $\varepsilon = 0.08$ is the coupling strength and $X(t) = \frac{1}{N}\sum_{k=1}^{N} x_k$ is the mean field. $\tau_0$ is a measure of internal time delay due to finite propagation speeds of signals, finite response time of synapses, etc., which having crucial role in collective dynamics of coupled neural oscillators [25]. Parameter $I_j$ is taken as $3.0 + \sigma$, where $\sigma$ is a Gaussian distributed number with zero mean and 0.01 rms value. Desynchronization in an ensemble (13) without coupling delay ($\tau_0 = 0$) has been studied with time-delayed feedback [15].

We have simulated the dynamics of an ensemble of $N = 2,000$ neurons with two different feedback strategies, namely washout-filter aided [17] and time-delayed feedback [15] (Fig. 4). In this example, the mean field does not vanish in the case of incoherent oscillations. It approaches to some constant instead [15] (Fig. 4a, $X_0 \approx -0.8294$, $t < 5,000$). Thus, to construct a vanishing control, we should get rid of the steady state of the mean field from the feedback [22]. Generally, the steady state is unknown practically, a washout-filter aided feedback strategy is applied, which results in robust desynchronization [17]. In the numerical example, we assume the collective signal $Y(t) = \frac{1}{N} \sum_{k=1}^{N} y_k$ is measured. Thus, the feedback system is of the following form

$$\begin{aligned} \dot{w}(t) &= Y(t) - d\, w(t) \\ v(t) &= Y(t) - d\, w(t) \end{aligned} \qquad (14)$$

where $w(t)$ and $v(t)$ are the state variable and output of the washout filter, respectively. The feedback is $u(t) = -g_w v(t)$ and is administered to all the oscillators through variables $y_j$, $g_w$ is the feedback gain. As shown in Fig. 4a, the mean field reduces to noise-level variations around $X_0$ when the feedback is set on (Fig. 4a, red line, $t > 10,000$). Moreover, there is little effort needed to maintain it (Fig. 4b, red line, $t > 1,000$). Because $u(t) \to 0$, $X \to X_0$, all the oscillators are subjected to a common force $\varepsilon X_0$, which however has little effect on their natural behaviors for weak coupling strength $\varepsilon$ (Comparing Fig. 4d with Fig. 4c). It is independent of the magnitude of feedback gain $g_w$.

In the numerical example, a time-delayed feedback is also used, which is of the form $u(t) = -g_\tau Y(t - \tau)$ [15], where $g_\tau$ is the feedback gain. Although the external time-delayed feedback may induce undesirable instability [15], it can suppress the large mean field oscillations with appropriate time-delay $\tau$ and feedback gain $g_\tau$, as shown in Fig. 4a. However, the feedback drives the mean field to the origin instead of its natural steady state $X_0$, thus resulting in a continuing input $g_\tau Y_0$ (Fig. 4b, blue line). The non-vanishing stimulations have significant impact both on the collective dynamics of the mean field (Fig. 4a, blue line, $t > 10,000$), and on the individual activities with large feedback gain $g_\tau$ (Comparing Fig. 4e with Fig. 4c). The larger the feedback gain $g_\tau$, the stronger the impact.

In conclusion, a general top-down approach is presented for desynchronization in large ensemble of weekly coupled oscillators. Dealing directly with global rhythmic behaviors, it should be independent of any details of the constituent elements as well as the type of their interactions and network topology of real systems. We anticipate our approach will have important applications in physical, biological, and ecological systems, etc, to destroy deleterious synchronization [12]. For example, it may help engineers to apply active or passive control methods to steady a footbridge so as to alleviate harmful crowd synchrony [9]. It may also find applications in ecological systems. In conservation ecology, synchronization is often perceived as detrimental because coherent oscillations of spatially structured subpopulations increase the danger of global extinction, while asynchrony enhances the global persistence of a population through "rescue effects" [26]. Since the underlying causes remain a long-standing enigma [27], it would be a big challenge to prevent the dangerous synchronization. We demonstrate that conservation measures, which balance the metapopulation as a whole, disrupt the coherence in subpopulation oscillations regardless of inherent synchronizing mechanisms. Hence, our work may help in developing effective global conservation policies to avoid global extinction [25]. The most relevant applications may be in neuroscience. For example, in neurological diseases such as Parkinson's disease, brain function is severely impaired by pathological synchrony in a group of neurons [10]. In these medical applications, it is desired to disrupt abnormal overall brain rhythms resulted from extremely strong neural synchrony, terminating its associated physiological effects [10, 13]. For the complexity of neurons and of neural networks, it would be impossible to desynchronize such a huge group of oscillatory neurons by taking into account of all these details. Our findings may provide an efficient way for this formidable task, namely by applying deep brain stimulation to suppress the resultant abnormal cortex rhythms registered by electrocorticogram (ECoG) [28]. Furthermore, a non-invasive stimulation is obtained by getting rid of the steady state of ECoG signals, which stops related pathological neural synchrony without altering the essential nature of neurons.

Acknowledgements: This work is supported by the State Key Program of National Natural

Science Foundation of China under Grant No. 11032009 , the Fundamental Research Funds for the Central Universities Shanghai and Leading Academic Discipline Project in No. B302.

[1] A. Pikosky, et al., *Synchronization, a Universal Concept in Nonlinear Sciences.* (Cambridge University Press, 2001).

[2] A. T. Winfree, *The Geometry of Biological Times.* (Springer, 1980).

[3] Y. Kuramoto, *Chemical Oscillations, Waves and Turbulence.* (Springer-Verlag, Berlin, 1984).

[4] S. M., Reppert, and D. R. Weaver. Nature 418, 935 (2002). D. Gonze, et al., Biophys. J. 89, 120 (2005).

[5] J. Buck, Quart. Rev. Biol. 13, 301 (1938).

[6] S. S. Wang and H. G. Winful, Appl. Phys. Lett. 52, 1774 (1988).

[7] K.-T. Kim, M.-S. Kim, and Y. Chong, Appl. Phys. Lett. 88, 062501 (2006).

[8] T. Danino, O. Mondragon-Palomino, L. Tsimring, and J. Hasty, Nature 463, 326 (2010).

[9] S. H. Strogatz, et al., Nature 438, 43 (2005); B. Eckhardt, E. Ott, S.H. Strogatz, D.M. Abrams, A. McRobie, Phys. Rev. E 75, 021110 (2007).

[10] W. W. Alberts, E. J. Wright, and B. Feinstein, Nature 221, 670 (1969).

[11] R. Abta, M. Schiffer, and N. M. Shnerb, Phys. Rev. Lett. 98, 098104 (2007); R. Abta, and N. M. Shnerb, Phys. Rev. E 75, 051914 (2007).

[12] W. L. Kath, and J. M. Ottino, Science 316, 1857 (2007).

[13] M. L. Kringelbach, N. Jenkinson, S. L. F. Owen, and T. Z. Aziz, Nature Rev. Neurosci. 8, 623 (2007).

[14] P. A. Tass, *Phase Resetting in Medicine and Biology: Stochastic Modeling and Data Analysis.* (Springer, Berlin, 1999).

[15] M. G. Rosenblum, and A. S. Pikovsky, Phys. Rev. Lett. 92, 114102 (2004); O. V. Popovych, C. Hauptmann, and P. A. Tass, Phys. Rev. Lett. 94, 164102 (2005); O. V. Popovych, C. Hauptmann, and P. A. Tass, Biol. Cybern. 95, 69 (2006).

[16] N. Tukhlina, M. Rosenblum, A. Pikovsky, J. Kurths, Phys. Rev. E, 75, 011918 (2007); N. Tukhlina, and M. G. Rosenblum, J. Biol. Phys. 34, 301 (2008).

[17] M. Luo, Y. J. Wu, and J. H. Peng, Biol. Cybern. 101, 242 (2009); M. Luo, and J. Xu, Neural Networks 24, 538 (2011).

[18] I. Z. Kiss, et al., Science 316, 1886 (2007).

[19] Y. Y. Liu, J. J. Slotine, and A.L. Barabasi, Nature 473, 167 (2011).

[20] D. G. Aronson, G. B. Ermentrout, and N. Koppel, Physica D 41, 403 (1990).

[21] R. E. Mirollo, and S. H. Strogatz, J. Stat. Phys. 60, 245 (1990).

[22] K. J. Astrom, and R. M. Murray, *Feedback Systems: an Introduction for Scientists and Engineers*. (Princeton University Press, 2008).

[23] P. C. Matthews, and S. H. Strogatz, Phys. Rev. Lett. 65, 1701 (1990).

[24] J. L. Hindmarsh, and R. M. Rose, Proc. R. Soc. London, Ser. B 221, 87 (1984).

[25] M. Dhamala, V. K. Jirsa, and M. Z. Ding, PHys. Rev. Lett. 92, 074104 (2004); E. Niebur, H.G. Schuster, and D.M. Kammen, Phys. Rev. Lett. 67, 2753 (1991); W. Gerstner, Phys. Rev. Lett. 76, 1755 (1996).

[26] M. Heino, V. Kaital, E. Ranta, and J. Lindstrom, Proc. R. Soc. Lond. B 264, 481 (1997); D. J. D. Earn, S. A. Levin, and P. Rohani, Science 290, 1360 (2000); R. E. Amritkar, and G. Rangarajan, Phys. Rev. Lett. 96, 258102 (2006).

[27] I. Hanski, Nature 396, 41 (1998); A. Liebhold, W. D. Koenig, and O. N. Bjornstad, Annu. Rev. Ecol. Evol. Syst. 35, 467 (2004).

[28] B. A. Lopour, A. J. Szeri, J. Comput. Neurosci. 28, 375 (2010); M. A. Kramer, B. A. Lopour, H. E. Kirsch, and A. J. Szeri, Phys. Rev. E 73, 041928 (2006); B. J. Gluckman, H. Nguyen, S. L. Weinstein,and S. J. Schiff, J. Neurosci. 21, 590 (2001).

Figure Captions

Fig. 1  Bifurcation diagrams of Eqs. (1) and (2) in $g-\Delta$ plane, $\varepsilon=0.5$ (a), and $g-\varepsilon$ plane, $\Delta=0.1$ (b). Regions I and III is phase locked region. Region II corresponding to incoherent region.

Fig. 2  Evolutions of the amplitude of the mean field $R$ and of individual units $\rho$ (a), and the phase difference $\phi$ with feedback gain $g$. The region $\varepsilon-\Delta\leq g\leq\varepsilon+\Delta$ corresponds to incoherent region.

Fig. 3  Feedback effects on an ensemble of Landau-Stuart oscillators (8) with the linear feedback (9) for $N=500$, $\varepsilon=0.5$, $g=-5.0$, and $\omega_j$ are random numbers uniformly selected in $[1-\gamma,1+\gamma]$ with $\gamma=0.05$. The coupling and feedback are set on at $t=500$ and $t=1,000$, respectively. (a) Time courses of the mean field $\text{Re}(Z)$ and feedback signal $|u|$. Time courses of two arbitrary oscillators in the uncoupled region (b) and the desynchronized region (c).

Fig. 4  Effects of vanishing and non-vanishing feedbacks on the collective behaviors and individual activities of neural ensemble (13). The internal coupling delay $\tau_0=0.015$. (a) Time courses of the mean field $X$ under washout-filter aided feedback for $d=1.0, g_w=5.0$ (red line) and time-delayed feedback for $\tau=0.2, g_\tau=5.0$ (blue line). The coupling and feedback are set on at $t=5,000$ and $t=10,000$, respectively. (b) Vanishing washout-filter aided feedback (red line) and non-vanishing time-delayed feedback (blue line). Time course of a randomly selected neuron in the coupling- and control-free region (c), under washout-filter aided feedback (d), and time-delayed feedback (e).

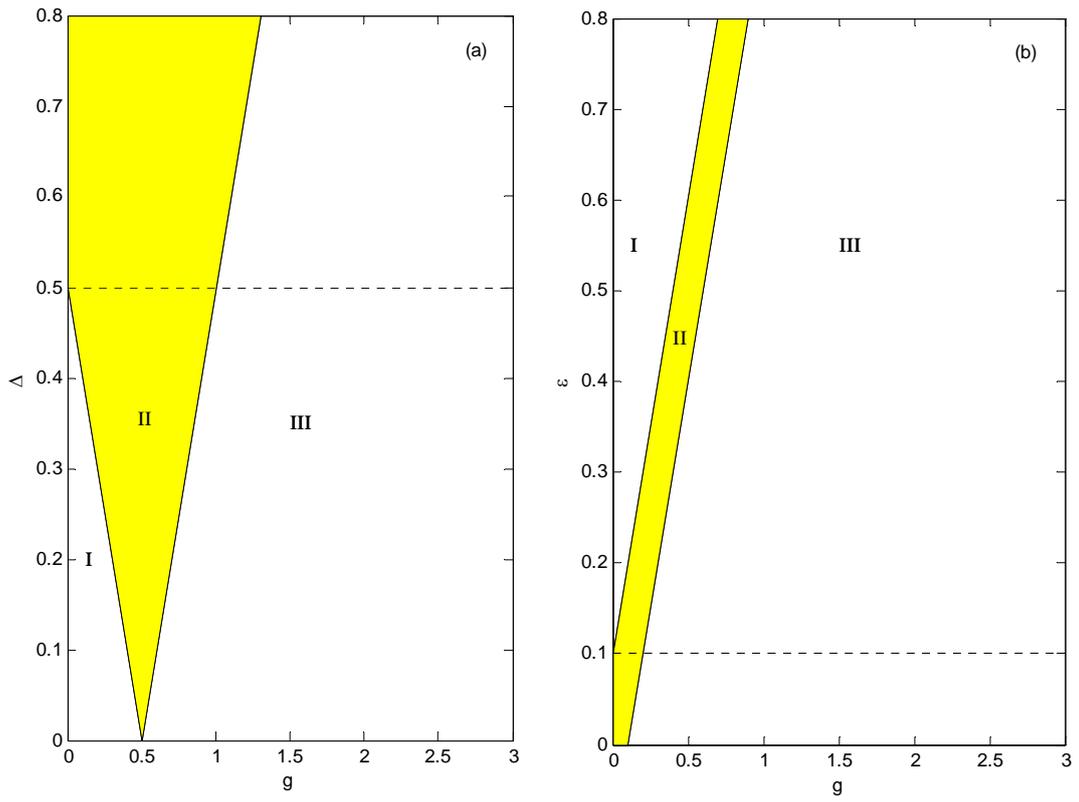

Fig. 1

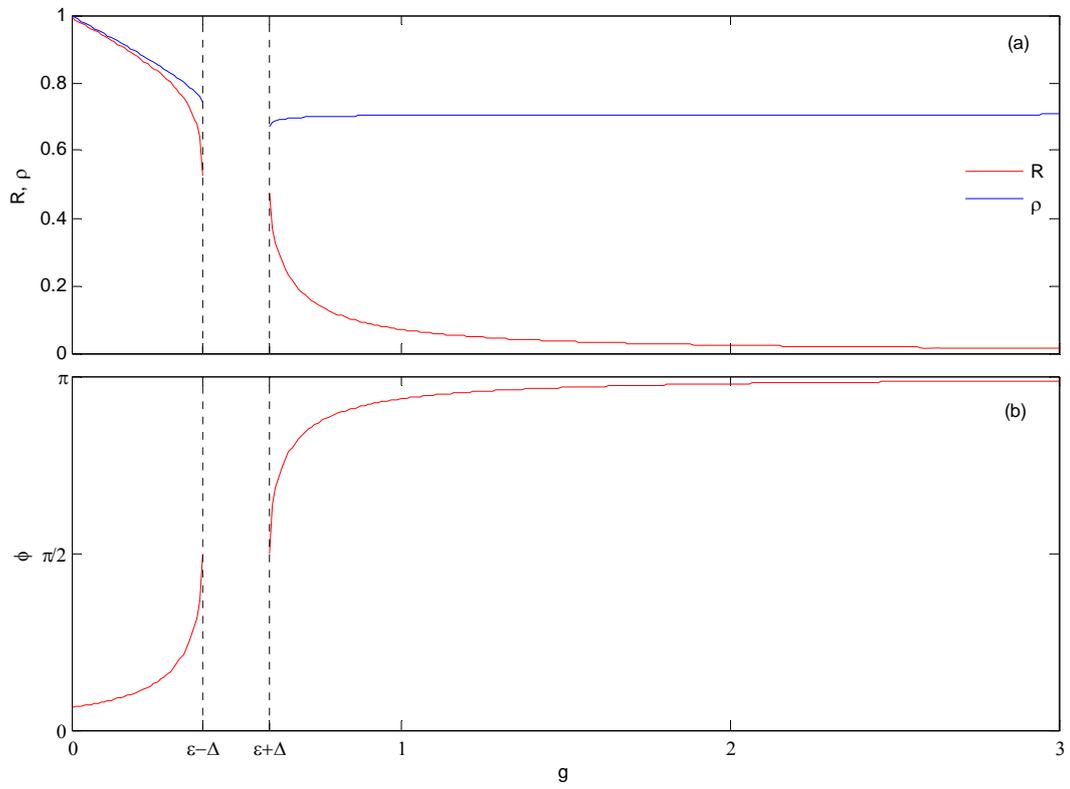

Fig. 2

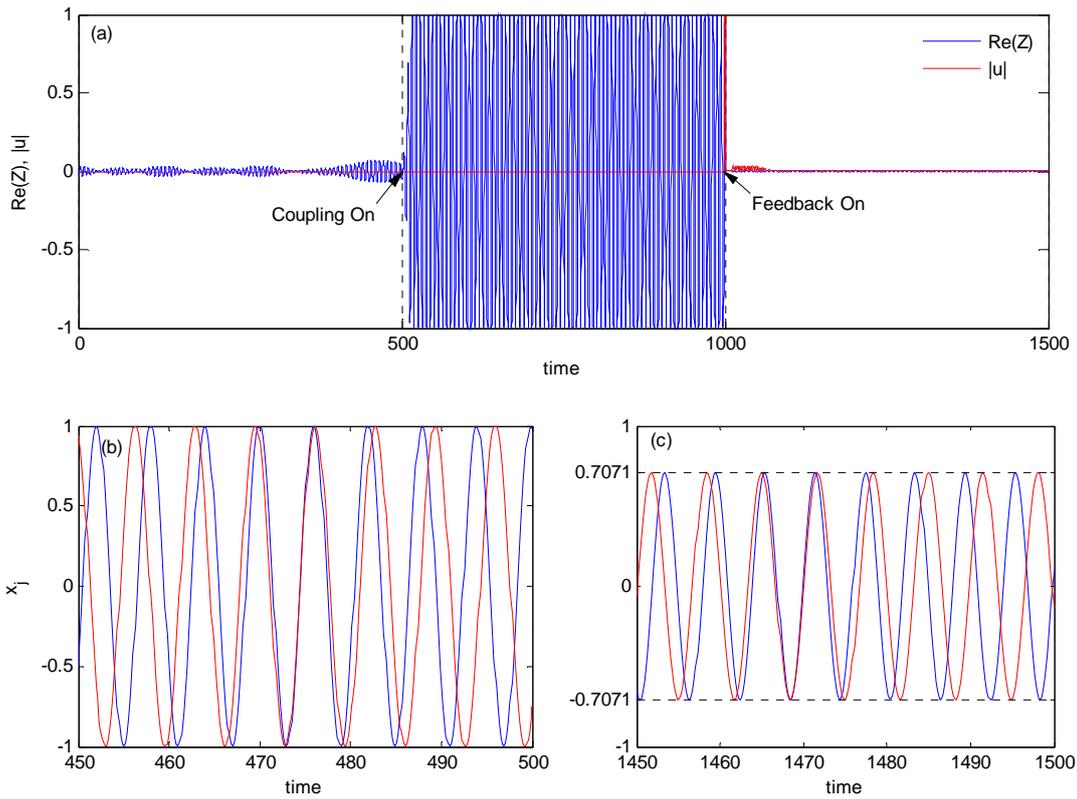

Fig. 3

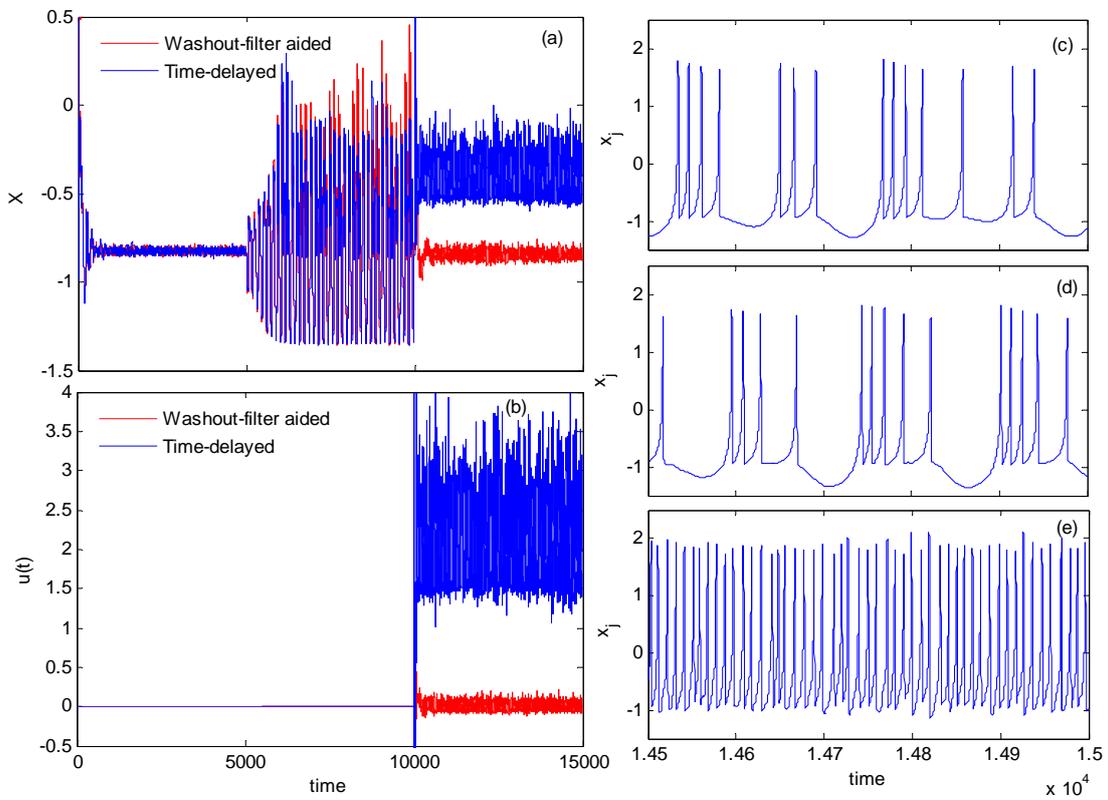

Fig. 4